\documentstyle[twocolumn,prl,aps,epsf,floats]{revtex}

\begin{document}
\draft
\def \beq{\begin{equation}}
\def \eeq{\end{equation}}
\def \beqarr{\begin{eqnarray}}
\def \eeqarr{\end{eqnarray}}

\twocolumn[\hsize\textwidth\columnwidth\hsize\csname @twocolumnfalse\endcsname

\title{ 
Reconstruction of Fractional Quantum Hall Edges
}

\author{Xin Wan$^1$, Kun Yang$^1$, and E. H. Rezayi$^2$}

\address{
$^1$National High Magnetic Field Laboratory and Department of Physics,
Florida State University, Tallahassee, Florida 32306
}

\address{
$^2$Department of Physics, California State University, Los Angeles,
California 90032}

\date{\today}
 
\maketitle

\begin{abstract}

We study the interplay of interaction, confining
potential and 
effects of finite temperature at the edge of a quantum Hall liquid.
Our exact 
diagonalization calculation  
indicates that edge reconstruction occurs in the fractional quantum
Hall regime for a variety of confining potential, including ones that 
correspond to a ``sharp" edge. 
Our finite temperature
Hartree-Fock calculation for integer quantum Hall edges 
indicates that reconstruction is suppressed above 
certain temperature. 
We discuss the implication of our results on recent edge
tunneling and microwave absorption experiments.

\end{abstract}

\pacs{
73.40.Hm, 71.10.Pm}
]

The physics of the quantum Hall edge states is important for a number of 
reasons. When the system is in a quantum Hall state, the bulk is incompressible
and the edge carries the dissipationless current. In addition,
the edge states also provide  
a unique 
arena to study correlated electrons in one-dimension, where the Fermi
liquid theory breaks down. Indeed, the edge electrons have been
argued\cite{wen}
to form the so-called chiral Luttinger liquid (CLL) in this
case, with
many properties determined solely by the robust bulk topological order
of the system,
and independent of the details of the edge potential. For example it was 
predicted\cite{wen}
that in the tunneling between a Fermi liquid and a quantum Hall edge, the
current-voltage relation follows a power-law $I\sim V^\alpha$, 
with universal exponents $\alpha$ for
simple edges like that of bulk filling $\nu=1/3$. 
While such power-law behavior 
has been observed in experiments indicating non-Fermi liquid 
behavior\cite{chang}, detailed studies of the bulk 
filling factor as well as sample dependence of $\alpha$
have found important differences from the predictions of CLL 
theory\cite{grayson,chang1,hilke}.

In the CLL theory\cite{wen}, one assumes that there is a single boundary 
separating the incompressible quantum Hall liquid from the electron vacuum,
and for simple edges like those of $\nu=1$ and $\nu=1/3$, the only low-energy
excitations are the shape fluctuations of the boundary, which form
a single branch of chiral bosons. 
On the other hand it is known that for integer quantum Hall edges,
the competition between electron-electron Coulomb interaction and edge 
confining potential can lead to edge reconstruction\cite{macdonald,chamon},
when the edge confining potential is smooth enough. At the reconstructed edge,
the electron density oscillates and there are more than one 
chiral boson modes, which do not propagate in the same direction\cite{chamon}.
Thus edge reconstruction can change the physics at the edge qualitatively.

Motivated by
the edge tunneling puzzle, as well as a recent microwave absorption
measurement in which the presence of artificial edge channels leads to
absorption enhancement\cite{ye}, we examine in this work the possibility of 
edge reconstruction when the bulk filling is fractional.
We find numerical evidence from exact diagonalization
studies suggesting that edge reconstruction
can occur when the bulk filling is fractional as well, for a variety of 
edge confining potentials (including those we believe properly describe a sharp
cleaved edge), and filling factors. 
We also study the effect of finite temperature on edge 
reconstruction using Hartree-Fock approximation on a $\nu=1$ edge, where we
find thermal fluctuations tend to suppress edge reconstruction. 
We argue that our results are highly relevant to and shed considerable light
on both the edge tunneling 
and the microwave absorption experiments.
We also comment on the differences and similarity between
our results and those of a recent numerical 
study of the $\nu=1/3$ edge\cite{goldman}. 

\begin{figure}
\epsfxsize=3.6in
\centerline{ \epsffile{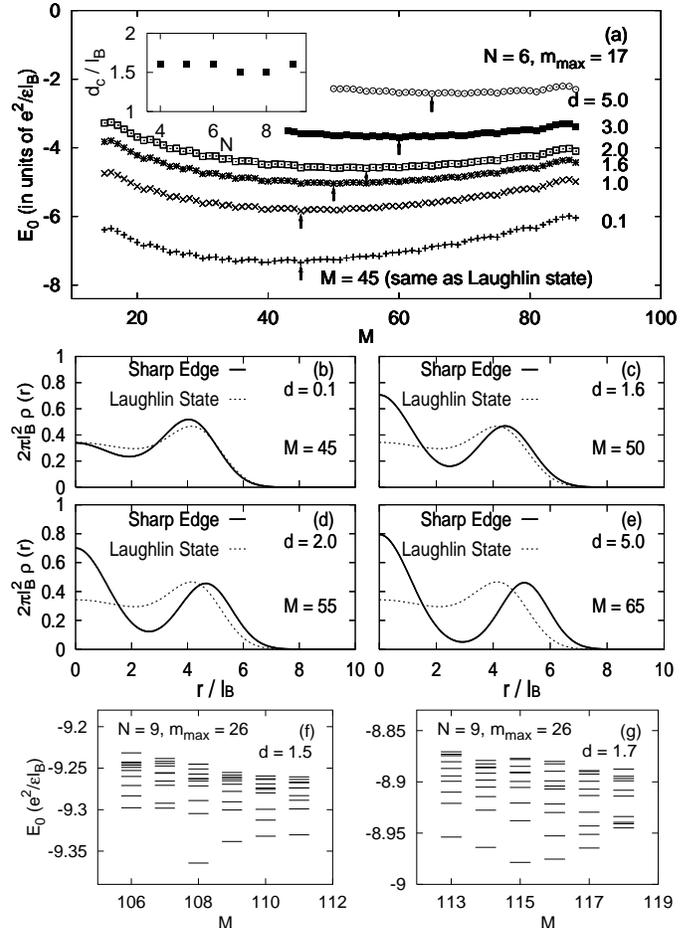} }
\caption{
\label{sharpEdge}
(a) Ground state energy
of each total angular momentum ($M$) subspace
as a function of
$M$ for 6 electrons in 18 orbitals,
with the background charge uniformly distributed on a disc above the
electron layer with a distance of $d$ from 0.1 to $5 l_B$.
The global ground state for each $d$ is indicated by an arrow below the
corresponding curve. The
inset shows the critical $d$ above which $M$ of the global ground state
differs from the corresponding
Laughlin state as a function of the number of electrons $N$ at $\nu=1/3$.
The electron densities $\rho(r)$
of the global ground state for 6 electrons
are compared with that of the Laughlin state (dotted lines)
for (b) $d = 0.1$, (c) $d = 1.6$, (d) $d = 2.0$ and (e) $d = 5.0$,
in units of $l_B$.
The low-lying states near the global ground state are shown for $N = 9$
electrons for (f) $d = 1.5$ (before edge reconstruction) and (g) $d = 1.7$
(after edge reconstruction).
}
\end{figure}

\begin{figure}
\epsfxsize=3.2in
\centerline{ \epsffile{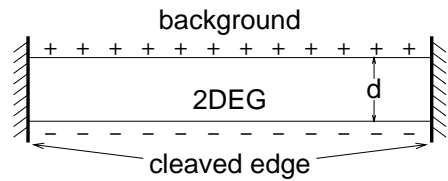} }
\caption{
\label{capacitor}
Side view of the electron gas layer and the background charge layer with
a sharp cleaved edge.
}
\end{figure}

{\em Exact Diagonalization Study} {--}{--}
In this work we diagonalize the following Hamiltonian exactly, which describes
electrons confined to the lowest Landau level, using the symmetric gauge:
\beq
H={1\over 2}\sum_{mnl}V_{mn}^l c_{m+l}^\dagger c_n^\dagger c_{n+l}c_m
+\sum_m U_mc_m^\dagger c_m,
\eeq
where $c_m^\dagger$ is the electron creation operator for the lowest Landau 
level single electron state with angular momentum $m$, $V_{mn}^l$ is the 
matrix element of Coulomb interaction, and $U_m$ is the matrix element of
a rotationally invariant confining potential
due to the positive background charge.
We have studied systems with $N=4$ to $12$ electrons, at a number of
filling factors.
Figure~\ref{sharpEdge}(a) shows the ground state energy\cite{backnote}
$E_0(M)$ of a given subspace  
with a fixed total angular momentum $M$ as a function of
$M$, for $N=6$ electrons.
The confining potential $U_m$ is calculated from a disc of uniformly
distributed 
positive background charge placed at a distance $d$ from the
electron layer, and the radius of the disc $a$ is chosen such that
the disc encloses 18 flux quanta.
The charge of the disc is always chosen to be the same as
that of the electrons so that the system is neutral.
Furthermore, we restrict the electrons to 18
orbitals, from $m = 0$ to $m_{max} = 17$; we believe this properly describes a
sharp cleaved edge where the electrons cannot move beyond where the confining
charge ends. Thus in this case
our model describes a 6 electron system at $\nu= 1/3$, 
confined by both the background charge and a sharp edge\cite{note}.
Fig.~\ref{capacitor} is an illustration of our model.
As we change the confining potential by increasing $d$ 
from $d = 0.1$ to 5,
in units of the magnetic length $l_B$, the total angular momentum $M$ 
of the global ground state increases from $M = 45$ to 65.
We note that the Laughlin state for 6 electrons has total angular momentum 
$M=45$. 
This indicates that while the global ground state of the system is
well described by the Laughlin state for $d < 1.6$, edge reconstruction occurs
for $d \ge 1.6$, as indicated by the increase of $M$\cite{macdonald,chamon}.

The presence of edge reconstruction can also be seen in the electron density
profile near the edge, as we plot 
in Fig.~\ref{sharpEdge}(b)-(e), and compare them with
the Laughlin state. 
Despite the Coulomb interaction and the confining potential, the
electron density for $d = 0.1$ closely resembles the Laughlin state,
with slightly stronger oscillation which is characteristic of the Coulomb
interaction.  
Our calculations show that the critical $d$ 
above which the total angular momentum $M$ of the global 
ground state differs from the
Laughlin state is between 1.5 and 1.6 for this system.
The corresponding electron density for $d = 1.6$
(Fig.~\ref{sharpEdge}(c)) shows much larger oscillation than that of
the Laughlin state. 
The critical $d$ fluctuates around $d_c = 1.6 \pm 0.1$ for system as
large as 9 electrons in 27 orbitals (Fig.~\ref{sharpEdge}(a) inset).
It is thus reasonable to expect
that the critical $d$ for edge reconstruction 
at $\nu = 1/3$ lies between 1 and $2 l_B$ 
in the thermodynamic limit, considerably
smaller than the typical $d$ in experiments ($5 - 10 l_B$). 
The change of $M$ at $d=d_c$ increases with $N$ as expected.
As $d$ further increases, the oscillation becomes stronger 
so that the electron density gradually separates into an edge piece 
and a center piece. 
The edge piece is almost detached from the center piece for $d = 5.0$, 
providing further evidence of edge reconstruction at fractional fillings. 

Edge reconstruction is the result of competition between Coulomb interaction
and confining potential\cite{macdonald,chamon}. In our model the confining 
potential is controled by $d$; the larger $d$ is, the weaker the confinement is
at the edge due to the backgroud charge, thus edge reconstruction occurs above
certain $d_c$. 
Another way to understand this fact is that without edge 
reconstruction, the electron density is approximately uniform all the
way to the edge, thus the electron gas layer and the background charge
layer form a capacitor (see Fig.~\ref{capacitor}). 
In such a capacitor the electrostatic potential is a constant in the
bulk of the electron layer, but a gradient (or fringe electric field with 
in-plane component) develops at the edge, which tends to pull the electrons 
toward the edge. The larger $d$ is, the larger the area is affected by this
fringe field, and eventually reconstruction occurs. Since in the systems we 
study the linear size of the system is much larger than $d_c$, the finite size
effects are weak (as also indicated by the weak dependence of $d_c$ on $N$); 
a detailed analysis of the finite size effects will be
presented elsewhere. We also note an important difference bewteen the
integer and fractional edge reconstruction; in the former the presence
of a sharp boundary alone eliminates the possibility of reconstruction
due to Pauli principle; it is no longer the case for the fractional edge.

Besides the single chiral mode predicted by the CLL theory\cite{wen}, 
additional edge modes are expected for reconstructed edges\cite{chamon}. We
find evidence for this in the spectra. Figs. 1(f) and 1(g) are the low-energy 
spectra for $N=9$ systems at $\nu=1/3$, before and after edge reconstruction.
We find before reconstruction low-lying states only exist for 
$M > M_{ground} =108$, consistent with the presence of a single chiral mode;
while after reconstruction there are low-lying states with $M$ both above and
below $M_{ground}=115$, indicating that there are modes propagating in both
directions.

\begin{figure}
\epsfxsize=3.6in
\centerline{ \epsffile{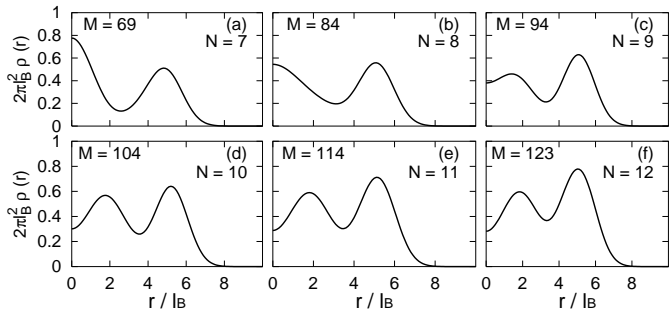} }
\caption{
\label{beyond6}
The electron density $\rho(r)$ of the lowest energy state for
$N = 7$ to 12 electrons in 18 orbitals with uniform background
charge, distributed $d = 3 l_B$ above the electron layer.
}
\end{figure}

Indication of edge reconstruction persists when we increase the average filling
factor from $\nu = 1/3$ to $2/3$. 
Figure~\ref{beyond6}(a)-(f) show the electron densities of the global ground
states for up to 12 electrons in 18 orbitals.  
The background charge is confined within the same radius $a$ as above, 
at a distance of $3 l_B$ above the electron layer, 
with appropriate charge density to ensure charge neutrality.
In all cases, we find indication of edge reconstruction as for $N = 6$. 
As $N$ increases, the electron density in the edge piece increases
accordingly, while its peak location remains almost unchanged. 
It should be noted however, that in these cases the filling factor for 
the bulk is less well-defined,
as the ``bulk'' area is too small and the electron density still
oscillates at $r = 0$. 
Nevertheless, we believe that the tendency for electron density to
reconstruct at the edge is a generic feature for appropriately confined 
electrons at any fractional filling. 

While a sharp cutoff (at $m_{max}$)
where the background charge ends properly describes
a cleaved edge, it is not present for other edges. 
Not surprisingly, we find that removing this
sharp cutoff further favor edge reconstruction. 
Evidence for edge reconstruction in a system with a smooth edge will be
presented elsewhere. 

\begin{figure}
\epsfxsize=3.6in
\centerline{ \epsffile{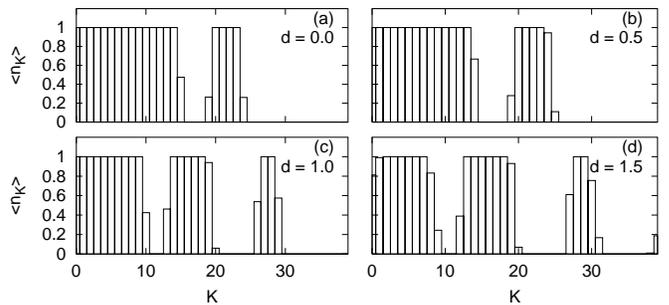} }
\caption{
\label{edgePattern}
Occupation numbers for background charge with $W = 10 l_B$ 
and different $d$ from 0 to $1.5 l_B$ calculated within finite-$T$ 
Hartree-Fock approximation at $T = 0.001e^2/\epsilon l_B$. 
}
\end{figure}

{\em Hartree-Fock Calculation} {--}{--}
We now turn to a discussion 
of the effect of a finite temperature $T$ on a reconstructed edge. 
To this end we will use a finite-temperature Hartree-Fock (HF)
approximation to study the $\nu=1$ edge.  
Laying the background charge right on top of the electron layer, Chamon and
Wen~\cite{chamon} demonstrated that a lump of electron density can move
$\sim 2 l_B$ away from the $\nu=1$ bulk for smooth enough background
charge distribution. In their model 
the smoothness of the edge potential is controlled by the width $W$ over which
the positive background charge density decreases from the bulk value
to zero. They have also shown that the results of HF approximation
agree well with those of exact diagonalization at $T=0$.
Here we use the finite-temperature Hartree-Fock approximation for the
interacting electron system, with cylindrical geometry similar to the one
studied by Chamon and Wen~\cite{chamon}, with an extra
parameter $d$ which is the distance the background charge lies above the
electron gas. In Fig.~\ref{edgePattern}(a)-(d) we show results of the
average electron occupation numbers at very low
temperature $T = 0.001$, in units of $e^2 /
\epsilon l_B$,
for 20 edge electrons in 40 single particle states on a strip of length
$L = 20 l_B$ with periodic boundary conditions, 
background charge $W = 10 l_B$, and $d$ varying from $d = 0$ to $1.5 l_B$.
We find that the edge reconstruction pattern depends on the detail of
the edge confining potential.
States with linear momentum $K < 0$ (``bulk'') and $K \geq 40$ are
assumed to be inert, with occupation number $n_K$ fixed to be 1 and 0,
respectively. 
Fig.~\ref{edgePattern}(a) essentially reproduces the $T=0$ results by
Chamon and Wen. 
As $d$ increases (so that the confining potential gets even smoother), 
the detached piece becomes wider and an extra piece appears for
$d = 1.0$ and 1.5. 
The size of the reconstructed region increases rapidly with $d$, and goes
beyond what we can handle numerically for $d \geq 2.0l_B$. 

\begin{figure}
\epsfxsize=3.6in
\centerline{ \epsffile{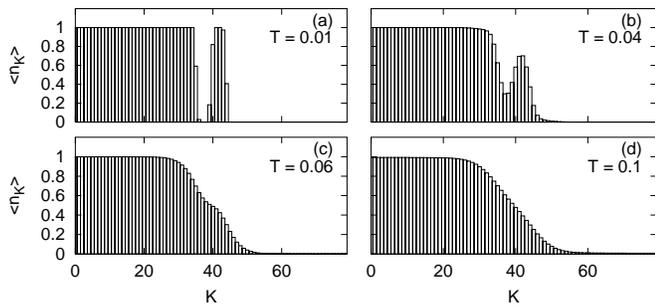} }
\caption{
\label{finiteT}
Occupation numbers for background charge with $W = 10 l_B$ 
and $d = 0$ at different $T$ within HF approximation. 
}
\end{figure}

The edge reconstruction disappears as we increase the temperature. 
Figure~\ref{finiteT}(a)-(d) show the occupation numbers for background
charge with $W = 10 l_B$ and $d = 0$ (Fig.~\ref{edgePattern}(a)) at higher temperatures. 
We double the number of active single particle states to 80, as well as
the number of edge electrons to $N = 40$, to ensure that 
the bulk states ($K < 0$) are
fully occupied (within 99\% percent), {\it a posteriori}
condition for separating the electrons into ``edge'' and ``bulk'' ones. 
The length of the strip remains at $L = 20 l_B$.
The detached edge piece at $T \leq 0.01$ mixes with the bulk at $T =
0.04$, although a well-defined peak can still be resolved.
The peak evolves to a shoulder at $T = 0.06$, disappearing before $T$
reaches 0.1, where the occupation number decreases smoothly from 1 to 0,
mirroring the smoothed background charge profile. 
Although extra pieces can emerge for finite $d$, we find that all edge
pieces merge with the bulk and disappear at $T \sim 0.05$.  
We thus conclude that, within the Hartree-Fock approximation, edge
reconstruction disappears above $T \sim 0.05 e^2 / \epsilon l_B$ at $\nu=1$. 
We have obtained a very similar temperature scale above which edge
reconstruction disappears for the $\nu=1/3$ edge, using results 
of exact diagonalization (that generates all eigenstates), 
in systems with up to 7 electrons; details
will be presented elsewhere.

{\em Discussions} {--}{--}
When
the edge reconstructs, there are multiple edge modes even for the simplest
quantum Hall states like $\nu=1$ and $\nu=1/3$\cite{chamon}. In particular,
these modes do not propagate in the same direction, and as a consequence the
tunneling exponent $\alpha$ depends on the interaction among these modes and
is thus no longer universal, even for these bulk fillings. Thus the edge
reconstruction can very well be the reason why $\alpha$ is always found to be
slightly less than the CLL prediction of $\alpha=3$ at $\nu=1/3$, 
and why it varies continuously with
bulk filling and other sample parameters, 
while CLL predicts plateau behavior\cite{theory}. 
On the other hand experiments find\cite{grayson} a very simple (but 
approximate) relation $\alpha\sim 1/\nu$ 
(see, however, Refs. \onlinecite{chang,hilke}),
for
which we do not have an explanation 
here. Nevertheless we believe edge reconstruction is likely to be an important
piece of the edge tunneling puzzle, and should be included as part of the
complete theory for the fractional quantum Hall edge. 

In a recent microwave absorption experiment\cite{ye}, it was found that
the conductivity at microwave frequency is strongly enhanced in samples 
with antidots in the fractional quantum Hall regime at very low temperatures,
while no such enhancement was found in similar samples with no antidots. It is
thus natural to associate the enhancement with the edge channels of the 
antidots. On the other hand, the dots are so small that it is 
easy to convince oneself that the frequency of the edge magneto plasma
(EMP) is much higher than the microwave frequency used in the
experiments (by about an order of magnitude). It thus appears unlikely
that the enhancement is  due to the microwave frequency approaching that
of EMP.  
On the other hand, a reconstructed edge can support softer acoustic
modes, which may  well be accessible in the experimental frequency
range\cite{sondhi}. 
Another important experimental finding is
that the enhancement disappears above temperature $T^*\sim 0.5 K$. 
This is qualitatively consistent with our finding that edge reconstruction
disappears at higher $T$. While the Hartree-Fock estimate of this
temperature scale: $T \sim 0.05 e^2 / \epsilon l_B\sim 6 K$, is higher
than $T^*$, we note that our calculations tend to overestimate this
temperature due to finite size effects as well as the fact that we have not
included
factors such as finite layer thickness, Landau level mixing and
disorder not included in our calculations; they all tend to lower this scale.
We thus conclude that the results of Ref. \onlinecite{ye} can be
explained as due to edge reconstruction at low temperatures. 

In a recent study, Tsiper and Goldman\cite{goldman} found electron density
oscillation near the edge of a electron droplet at $\nu=1/3$, which they 
interpret as formation of edge density wave. In their study the confining 
potential due to the background charge is totally neglected, while we believe 
it is extremely important for a proper understanding of the physics at the
edge. Moreover, they have focused only on states that have the same quantum 
number as the Laughlin state; as we demonstrated this is often {\em not} 
the case for the global ground state, a fact
which is precisely the indication of edge reconstruction. On the other hand 
the edge density oscillation they find is quite similar to those of 
reconstructed edges reported here.

The authors thank Lloyd Engel and Dan Tsui for stimulating discussions 
that motivated
the present work and useful comments on the manuscript, 
and Claudio Chamon, Allan MacDonald, and Shivaji Sondhi
for helpful conversations. 
EHR gratefully acknowledges the hospitality of the NHMFL 
where part of this work was performed. This work was supported by 
NSF grants DMR-9971541 (XW and KY), DMR-0086191 (EHR), 
the State of Florida (XW),
and the Sloan Foundation (KY).


\begin{references}
\bibitem{backnote} We do not include the self-interaction of the background 
charge (which is a constant for a given system size)
in our definition of the energy.
\bibitem{note} We note that for the Laughlin state at $\nu=1/3$, single 
electron states are populated only up to $m_L=3(N-1)=m_{max}-2=15$ 
for $N=6$. We have 
checked that choosing $m_{max}$ to be $m_L$ instead
makes very small differences to $d_c$, and the difference shrinks to zero
as $N$ increases.
\end{references}
\end{document}